\documentclass[prl,twocolumn,superscriptaddress]{revtex4}
\usepackage{amsmath,graphicx}
\usepackage{epsfig}
\usepackage{paralist}
\usepackage{amsmath,amssymb,mathrsfs}
\usepackage{underscore}
\usepackage{natbib}
\usepackage{color}
\usepackage{dsfont}
\usepackage{bm}
\usepackage[colorlinks,linkcolor=blue,anchorcolor=blue,urlcolor=blue,citecolor=blue]{hyperref}

\begin{document}

\title{Higher-order topological insulator with glide reflection symmetry}

\author{Shaoliang Zhang}
\email{shaoliang@hust.edu.cn}
\affiliation{School of Physics, Institute of Quantum Science and Engineering, Huazhong University of Science and Technology, Wuhan 430074, China}

\date{\today}

\begin{abstract}

The Benalcazar-Bernevig-Hughes(BBH) model is one of the most representative higher-order topological insulators with different bulk-edge correspondence comparing conventional topological states. In this work, we propose to realize a couple-BBH model with glide reflection symmetry by coupling two topologically distinct BBH models. Such a system gives rise to intriguing properties beyond the traditional BBH model. A topological phase transition from higher-order topological phase to a semimetal has been predicted in this system. Furthermore, the band crossing points and nodal ring protected by glide reflection symmetry has also been explored. Our work provides a viewpoint to demonstrate the versatile phenomena induced by the interplay between higher-order topology and nonsymmorphic symmetries. 

\end{abstract}

\maketitle

\section{introduction}
\label{sec1}

Two-dimensional Benalcazar-Bernevig-Hughes(BBH) model\cite{Bernevig1,Bernevig2}, as one of the most representative model of higher-order topological insulator(HOTI), has attracted a lot of interests in condensed matter physics because of its simple structure and beautiful topological properties. Different with conventional topological states\cite{mzhasan}, there is no gapless edge state when BBH model is in topological non-trivial phase no matter which direction is chosen as open boundary. But if the open boundary is chosen as a square, four topologically protected zero-energy corner states emerge on the corners of the system. This new type of bulk-edge correspondence not only extends the understanding of topological physics\cite{trifunovic}, but also inspires new viewpoints to the application of topological matters\cite{banerjee,cli}. Recently, the research about HOTI developed rapidly and a series of theoretical models about HOTI have been proposed\cite{hotimodel1, hotimodel2, hotimodel3, hotimodel4, hotimodel5, hotimodel6, hotimodel7}. Some of them have been constructed experimentally in some synthetic quantum systems and metamaterials\cite{metamaterial1, metamaterial2, metamaterial3, metamaterial4, metamaterial5, metamaterial6}, and the gapless corner states has been observed. The HOTI has also been observed in some condensed matter systems\cite{hotiinbismuth1,hotiinbismuth2}.

The topology of higher-order topological states always depend on the symmetries\cite{symmetry1, symmetry2, symmetry3, symmetry4, symmetry5, symmetry6, symmetry7}. For example, the bulk topological properties of BBH model can be characterized by the quantized quadrupole moment, where the quantization is protected by the chiral symmetry and $C_2$ symmetry\cite{Bernevig1, Bernevig2}. In the previous works, these symmetries are always belonged to the point-group symmetries. In the research of symmetry protected topological states, people also focus on the role of space-group symmetries, especially the nonsymmorphic symmetries, which are the combination of point-group symmetric operations with fractional translations of lattice. The nonsymmorphic symmetry can lead to a lot of novel topological states because it will induce additional band crossing, such as nodal points or nodal lines\cite{glide1, glide2, glide3, glide4, glide5, glide6, glide7, glide8}. So far, there are comprehensive and thorough researches about nonsymmorphic symmetry protected topological states, but only a few of them involve the interplay between higher-order topology and nonsymmorphic symmetries\cite{glidehoti1, glidehoti2, glidehoti3}.

In this work, we propose a coupled-BBH model which can be constructed in ultracold atomic systems or some other synthetic quantum systems. It is constituted by two BBH models coupled with each other. We consider the ultracold atoms with two hyperfine spin states and each spin component is confined in a spin-dependent two-dimensional lattice respectively, where the lattice structure of different spin components has a fractional lattice distance. By fine tuning the form of coupling between different spin components, we can ensure the system has the glide reflection symmetry, which is one of the nonsymmorphic symmetries. We found that, although the system is a HOTI,  the topological properties depend on the amplitude of coupling instead of the ratio of intra and inter-cell tunneling in BBH model. Furthermore, the  glide reflection symmetry has significant effects to the properties of system. (1) If the coupling term preserves the glide reflection symmetry and time-reversal symmetry, and they commute with each other, the band structure has two-fold degeneracy and two degenerate states has opposite eigenvalues of glide operator respectively. (2) If the coupling term preserves the glide reflection symmetry but breaks time-reversal symmetry, the two-fold degeneracy will be lifted. But the glide reflection symmetry leads to additional band crossing and changes the topological properties of system. With the increase of coupling amplitude, the system changes from a HOTI to a semimetal. (3) If the coupling term breaks glide reflection symmetry and time-reversal symmetry individually but preserves their combination, which we call it preserves the magnetic glide symmetry, the symmetry protected nodal ring emerges in the system and the topological phase transition is also very different with above two cases.

This work is organized as follows. In Section \ref{sec2}, we introduce our model with on-site inter-spin coupling and discuss its topological properties. In Section \ref{sec3} and \ref{sec4}, we consider two different types of inter-spin couplings, one breaks the time-reversal symmetry but preserves glide reflection symmetry; the other breaks both but preserves their combination, i.e., the magnetic glide symmetry. In Section \ref{sec5}, the experimental setup to realize the discussed models in ultracold atomic systems has been proposed. Section \ref{sec6} is devoted to a discussion.

\section{Model and topological properties}
\label{sec2}

\begin{figure}[tph]
\centering
\includegraphics[width=76mm]{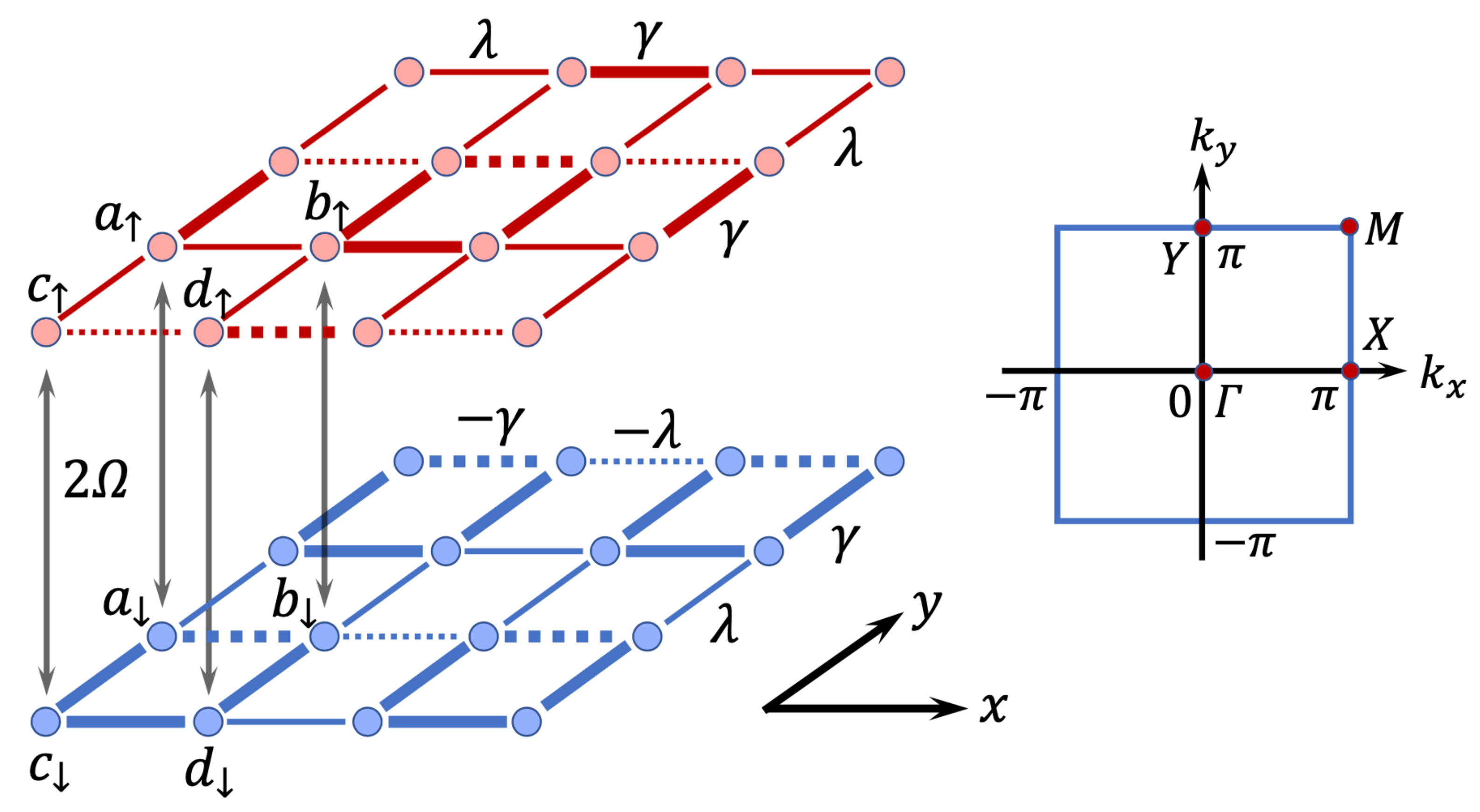}
\caption{The scheme of lattice structure. Red and blue dots correspond to different spin components. Thick and thin lines correspond to the amplitude of nearest-neighbor tunneling $\lambda$ and $\gamma$ respectively. Dotted line corresponds to the tunneling amplitude has an additional minus sign. Black double arrow lines correspond to the on-site coupling between different spin components. The inset on right shows the first Brillouin zone and some high symmetric points.} \label{fig:latticeonsitecoupling}
\end{figure}

We consider a square lattice with two different spin components. The lattice structure of each spin component is similar with BBH model\cite{Bernevig1, Bernevig2}, as shown in fig.\ref{fig:latticeonsitecoupling}, where the lattice structure for different spin components has a displacement along the diagonal line of $x$ and $y$ directions. An on-site coupling between different spin components has been added  and the tight-binding Hamiltonian can be written as
\begin{equation}
H=H_\uparrow+H_\downarrow+H_\mathrm{couple}
\label{hotiwithglide}
\end{equation}
where $H_\uparrow$ and $H_\downarrow$ are the Hamiltonian of spin-up and spin-down component respectively as 
\begin{equation}
\begin{split}
H_\uparrow=&\lambda\sum_{mn}\big(a^\dag_{mn,\uparrow}b_{mn,\uparrow}-c^\dag_{mn,\uparrow}d_{mn,\uparrow}\big) \\
+&\gamma\sum_{mn}\big(b^\dag_{mn,\uparrow}a_{(m+1)n,\uparrow}-d^\dag_{mn,\uparrow}c_{(m+1)n,\uparrow}\big) \\
+&\lambda\sum_{mn}\big(a^\dag_{mn,\uparrow}c_{mn,\uparrow}+b^\dag_{mn,\uparrow}d_{mn,\uparrow}\big) \\
+&\gamma\sum_{mn}\big(c^\dag_{mn,\uparrow}a_{(m+1)n,\uparrow}+d^\dag_{mn,\uparrow}b_{(m+1)n,\uparrow}\big)+h.c.
\end{split}
\label{bbhmodelup}
\end{equation}
\begin{equation}
\begin{split}
H_\downarrow=&\gamma\sum_{mn}\big(c^\dag_{mn,\downarrow}d_{mn,\downarrow}-a^\dag_{mn,\downarrow}b_{mn,\downarrow}\big) \\
+&\lambda\sum_{mn}\big(d^\dag_{mn,\downarrow}c_{(m+1)n,\downarrow}-b^\dag_{mn,\downarrow}a_{(m+1)n,\downarrow}\big) \\
+&\gamma\sum_{mn}\big(a^\dag_{mn,\downarrow}c_{mn,\downarrow}+b^\dag_{mn,\downarrow}d_{mn,\downarrow}\big) \\
+&\lambda\sum_{mn}\big(c^\dag_{mn,\downarrow}a_{(m+1)n,\downarrow}+d^\dag_{mn,\downarrow}b_{(m+1)n,\downarrow}\big)+h.c.
\end{split}
\label{bbhmodeldown}
\end{equation}
where $\lambda (\gamma)$ and $\gamma (\lambda)$ are amplitudes of intra- and inter-cell tunneling for spin-up (down) component respectively. $H_\mathrm{couple}$ is the coupling between different spin components as
\begin{equation}
\begin{split}
H_\mathrm{couple}=&2\Omega\sum_{mn}\big(a^\dag_{mn,\uparrow}a_{mn,\downarrow}+b^\dag_{mn,\uparrow}b_{mn,\downarrow} \\
&+c^\dag_{mn,\uparrow}c_{mn,\downarrow}+d^\dag_{mn,\uparrow}d_{mn,\downarrow}\big)+h.c.
\end{split}
\end{equation}
where $2\Omega$ corresponding to the coupling amplitude.  

When $\Omega=0$, different spin components are independent and the topological properties of the system is clear. If $\lambda<\gamma$, Hamiltonian $H_\uparrow$ is topological non-trivial but Hamiltonian $H_\downarrow$ is topological trivial. If $\lambda>\gamma$, the answer is opposite. So no matter in which case, the whole system is topological non-trivial. With a square open boundary, there will always be a zero energy state localized on each corner. In this work, we will only consider the $\lambda<\gamma$ case and set the length of square unit cell as $1$ for simplicity.

When $\Omega\ne 0$, the Hamiltonian (\ref{hotiwithglide}) is invariant with combination of spin flip $\uparrow\leftrightarrow\downarrow$ and the translation of the distance $1/\sqrt{2}$ along $x\pm y$ directions, which means the system has the glide reflection symmetry. With Fourier transformation, the above Hamiltonian (\ref{hotiwithglide}) can be rewritten in momentum space as $H=\sum_{\bf k}\Psi^\dag_{\bf k}H_{\bf k}\Psi_{\bf k}$ where $\Psi_{\bf k}=[a_{{\bf k},\uparrow},b_{{\bf k},\uparrow},c_{{\bf k},\uparrow},d_{{\bf k},\uparrow},a_{{\bf k},\downarrow},b_{{\bf k},\downarrow},c_{{\bf k},\downarrow},d_{{\bf k},\downarrow}]^\mathrm{T}$ and
\begin{equation}
\begin{split}
H_{\bf k}=&-\Gamma(1-\cos k_x)\tau_z\sigma_x+\Delta(1+\cos k_x)s_z\tau_z\sigma_x \\
&+\Gamma\sin k_x\tau_z\sigma_y+\Delta\sin k_xs_z\tau_z\sigma_y \\
&+\Delta(1+\cos k_y)\tau_x-\Gamma(1-\cos k_y)s_z\tau_x \\
&+\Delta\sin k_y\tau_y+\Gamma\sin k_ys_z\tau_y+\Omega s_x
\end{split}
\label{hotiwithglideinmomentumspace}
\end{equation}
where $\Gamma=(\lambda-\gamma)/2$ and $\Delta=(\lambda+\gamma)/2$. $\tau_i$, $\sigma_i$ and $s_i$ ($i=x,y,z$) are Pauli matrices which corresponding to orbital ($\boldsymbol{\tau}$, $\boldsymbol{\sigma}$) and spin ($\boldsymbol{s}$) degrees of freedom respectively. It can be separated into two parts as $H_{\bf k}=H_{d,{\bf k}}+H_\mathrm{couple}$. The block diagonalized matrix $H_{d,{\bf k}}=\mathrm{diag}\big[H_{{\bf k},\uparrow},H_{{\bf k},\downarrow}\big]$ corresponds to Hamiltonian of two different spin components where $H_{{\bf k},\uparrow}$ and $H_{{\bf k},\downarrow}$ are Hamiltonian of spin-up and spin-down components respectively, which can be expressed separately as
\begin{equation}
\begin{split}
H_{{\bf k},\uparrow}=&\big(\lambda+\gamma\cos k_x\big)\tau_z\sigma_x+\gamma\sin k_x\tau_z\sigma_y \\
&+\big(\lambda+\gamma\cos k_y\big)\tau_x+\gamma\sin k_y\tau_y
\end{split}
\end{equation}
\begin{equation}
\begin{split}
H_{{\bf k},\downarrow}=&-\big(\gamma+\lambda\cos k_x\big)\tau_z\sigma_x-\lambda\sin k_x\tau_z\sigma_y \\
&+\big(\gamma+\lambda\cos k_y\big)\tau_x+\lambda\sin k_y\tau_y
\end{split}
\end{equation}
$H_\mathrm{couple}=\Omega s_x$ corresponds to the coupling between different spin components. The glide operator can also be written in momentum space as $G_{\bf k}=s_x\tilde{G}_{\bf k}$ where 
\begin{equation}
\tilde{G}_{\bf k}=\left(\begin{array}{cc} 0 & 1 \\ e^{ik_y} & 0 \end{array}\right)\otimes\left(\begin{array}{cc} 0 & 1 \\ e^{ik_x} & 0 \end{array}\right)
\end{equation}
Here we just consider the glide operation along $x+y$ direction and the corresponding operation along $x-y$ direction also has the similar result. It can be verified that the Hamiltonian (\ref{hotiwithglideinmomentumspace}) satisfies the relation
\begin{equation}
G^\dag_{\bf k}H_{\bf k}G_{\bf k}=H_{\bf k}
\end{equation}
For each block of matrix $H_{d,{\bf k}}$, we also have the relation 
\begin{equation}
\tilde{G}^\dag_{\bf k}H_{{\bf k},\uparrow}\tilde{G}_{\bf k}=H_{{\bf k},\downarrow}
\label{blockglideops}
\end{equation}
Two consecutive glide operations correspond to one lattice translation along $x+y$ direction as
\begin{equation}
G^2_{\bf k}=e^{i(k_x+k_y)}\mathcal{I}
\label{eq:glidetwice}
\end{equation}
where $\mathcal{I}$ is identity matrix, which means that the eigenvalues of glide operator should be $\pm e^{i(k_x+k_y)/2}$.

\begin{figure}[tph]
\centering
\includegraphics[width=86mm]{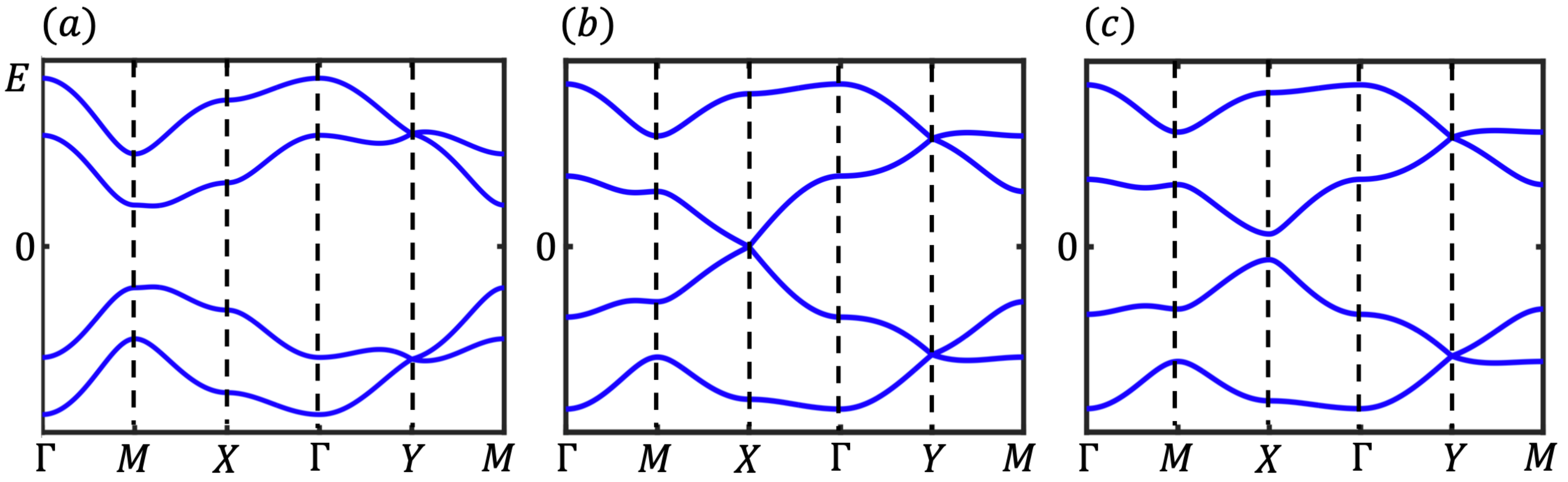}
\caption{The band structure of system with on-site coupling when $\lambda/\gamma=0.4$ for different coupling amplitude. (a) $\Omega/\gamma=0.4$, (b) $\Omega/\gamma=0.76$, (c) $\Omega/\gamma=0.9$. } \label{fig:bandOnsitecoupling}
\end{figure}

The band structure is shown in fig.\ref{fig:bandOnsitecoupling}. The system has time-reversal symmetry and chiral symmetry with symmetric operator as $\mathcal{T}=\mathcal{K}$ and $\mathcal{C}=s_z\tau_z\sigma_z$, where $\mathcal{K}$ corresponds to the Hermitian conjugation. Similar with BBH model\cite{Bernevig1, Bernevig2}, the system (\ref{hotiwithglideinmomentumspace}) has reflection symmetry along $x$ and $y$ directions with symmetric operator as $M_x=\sigma_x$ and $M_y=\tau_x\sigma_z$ respectively. It also has $C_2$ rotation symmetry with symmetric operator as $C_2=\tau_x\sigma_y$, which means that $C^{-1}_2H_{(-k_x,-k_y)}C_2=H_{(k_x,k_y)}$. These symmetries protect the quantization of quadrupole moment and our system is still a second-order topological insulator. But the topological properties of the Hamiltonian (\ref{hotiwithglideinmomentumspace}) depends on the coupling amplitude between different spin components. With the increase of $\Omega$, the system becomes from topological non-trivial phase to topological trivial phase. As shown in fig.\ref{fig:bandOnsitecoupling}(b), the critical point of topological phase transition is at
\begin{equation}
\Omega_\mathrm{cri}=\sqrt{\frac{\lambda^2+\gamma^2}{2}}
\end{equation}

Numerical calculation shows that there are four zero-energy corner states with square open boundaries when $\Omega<\Omega_\mathrm{cri}$. In this case, if we consider the bulk topology, four Wannier bands are all gapped and the corresponding Wannier sector polarization are all quantized and two of them are non-trivial and the other two are trivial\cite{Bernevig1, Bernevig2}. After passing through the critical point as $\Omega>\Omega_\mathrm{cri}$, the Wannier sector polarizations of all four Wannier bands are trivial and there is no zero-energy corner states with square open boundaries anymore.

In the whole range of the coupling amplitude $\Omega$, there is four-fold degeneracy at $Y=(0,\pi)$ point, as shown in fig.\ref{fig:bandOnsitecoupling}. We should point out that the degeneracy comes from the form of Hamiltonian (\ref{hotiwithglideinmomentumspace}) at $Y$ point only contains three terms $s_z\tau_z\sigma_x$, $s_z\tau_x$ and $s_x$, whose square are all identity matrix and anti-commute with each other. It doesn't relate to the glide reflection symmetry because this degeneracy can be lifted by adding an additional energy detuning $\delta\tau_z\sigma_z$ which preserve the glide reflection symmetry. Actually, the time-reversal symmetry and glide reflection symmetry commute with each other as $\big[\mathcal{T},G_{\bf k}\big]=0$ in this system. We can find an operator $\Theta=i\tau_x\sigma_y\mathcal{K}$ which is commute with the Hamiltonian (\ref{hotiwithglideinmomentumspace}) and preserves the Kramers degeneracy in the whole Brillouin zone(BZ) due to $\Theta^2=-1$. This operator and glide operator $G_{\bf k}$ also has the relation as
\begin{equation}
\Theta^{-1}G_{\bf k}\Theta=-e^{-i(k_x+k_y)}G_{\bf k}
\end{equation}
which means that if state $|\psi\rangle$ is the common eigenstate of Hamiltonian (\ref{hotiwithglideinmomentumspace}) and glide operator $G_{\bf k}$ with eigenvalue $E$ and $e^{i(k_x+k_y)/2}$ respectively, then $\Theta|\psi\rangle$ is degenerate with $|\psi\rangle$ and also the eigenstate of $G_{\bf k}$ with opposite eigenvalue as $-e^{i(k_x+k_y)/2}$. It implies that the eigenstates of glide operator with opposite eigenvalues always degenerate with each other in the whole BZ.

\section{The breaking of time-reversal symmetry}
\label{sec3}

\begin{figure}[tph]
\centering
\includegraphics[width=84mm]{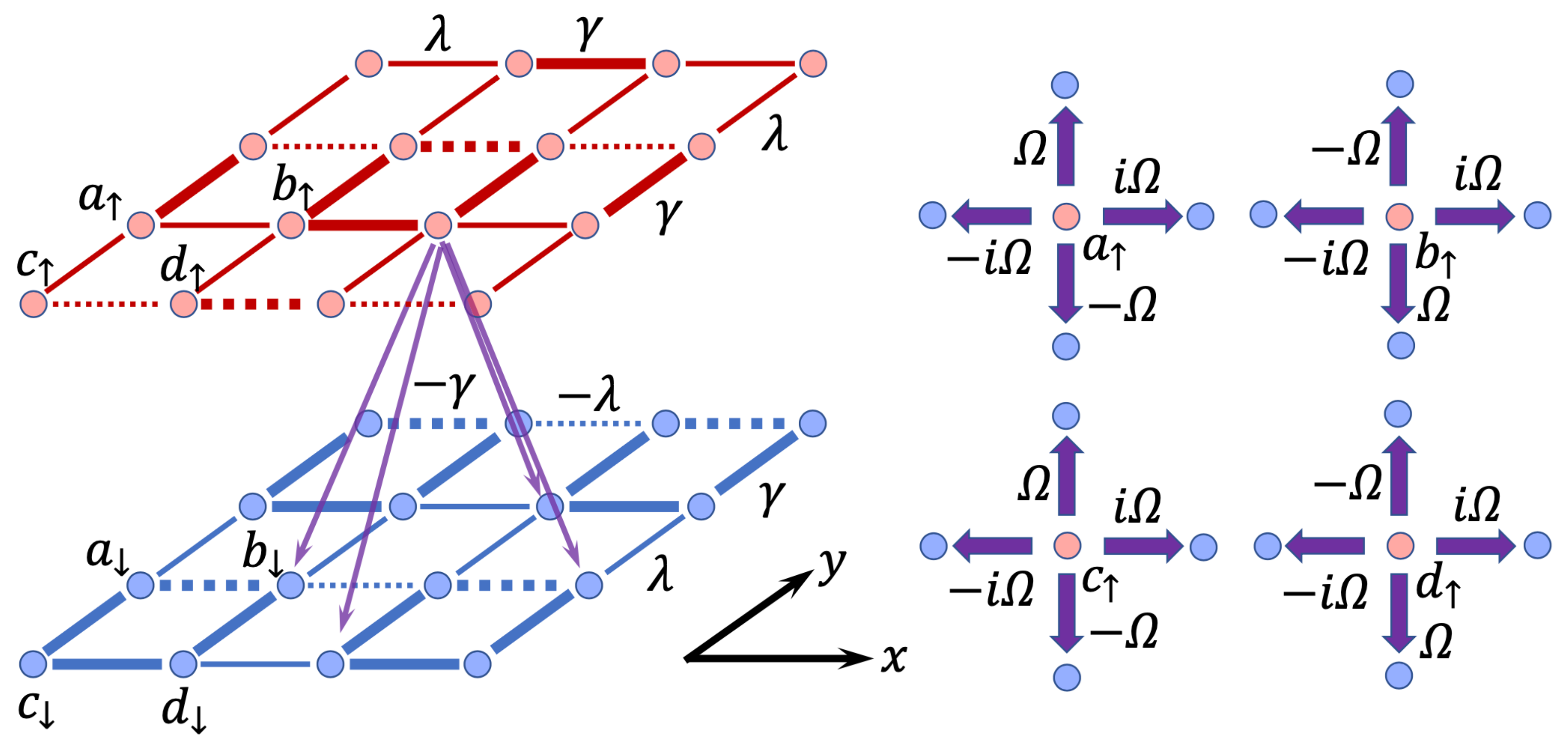}
\caption{Left: The scheme of lattice structure with nearest-neighbor coupling. Purple arrow lines correspond to the nearest-neighbor coupling between different spin components. Right: The tunneling from a spin-up site to its nearest-neighbor spin-down sites along different directions with different additional phases.} \label{fig:latticechiralcoupling}
\end{figure}

Instead of the on-site coupling between different spin components, one can add nearest-neighbor tunneling with additional phases along clockwise ($a_\uparrow$,$c_\uparrow$) and anti-clockwise ($b_\uparrow$,$d_\uparrow$) directions respectively, as shown in fig.\ref{fig:latticechiralcoupling}, which breaking the time-reversal symmetry. The Hamiltonian of two different spin components $H_{d,{\bf k}}$ has the same form with (\ref{hotiwithglideinmomentumspace}) and the coupling term changes into 
\begin{equation}
\begin{split}
H_\mathrm{couple,{\bf k}}&=\Omega(1-\cos k_x)s_x\sigma_y+\Omega\sin k_xs_x\sigma_x \\
+&\Omega(1-\cos k_y)s_y\tau_y\sigma_z+\Omega\sin k_ys_y\tau_x\sigma_z
\label{hotiwithchiralcoupling}
\end{split}
\end{equation}

\begin{figure}[tph]
\centering
\includegraphics[width=86mm]{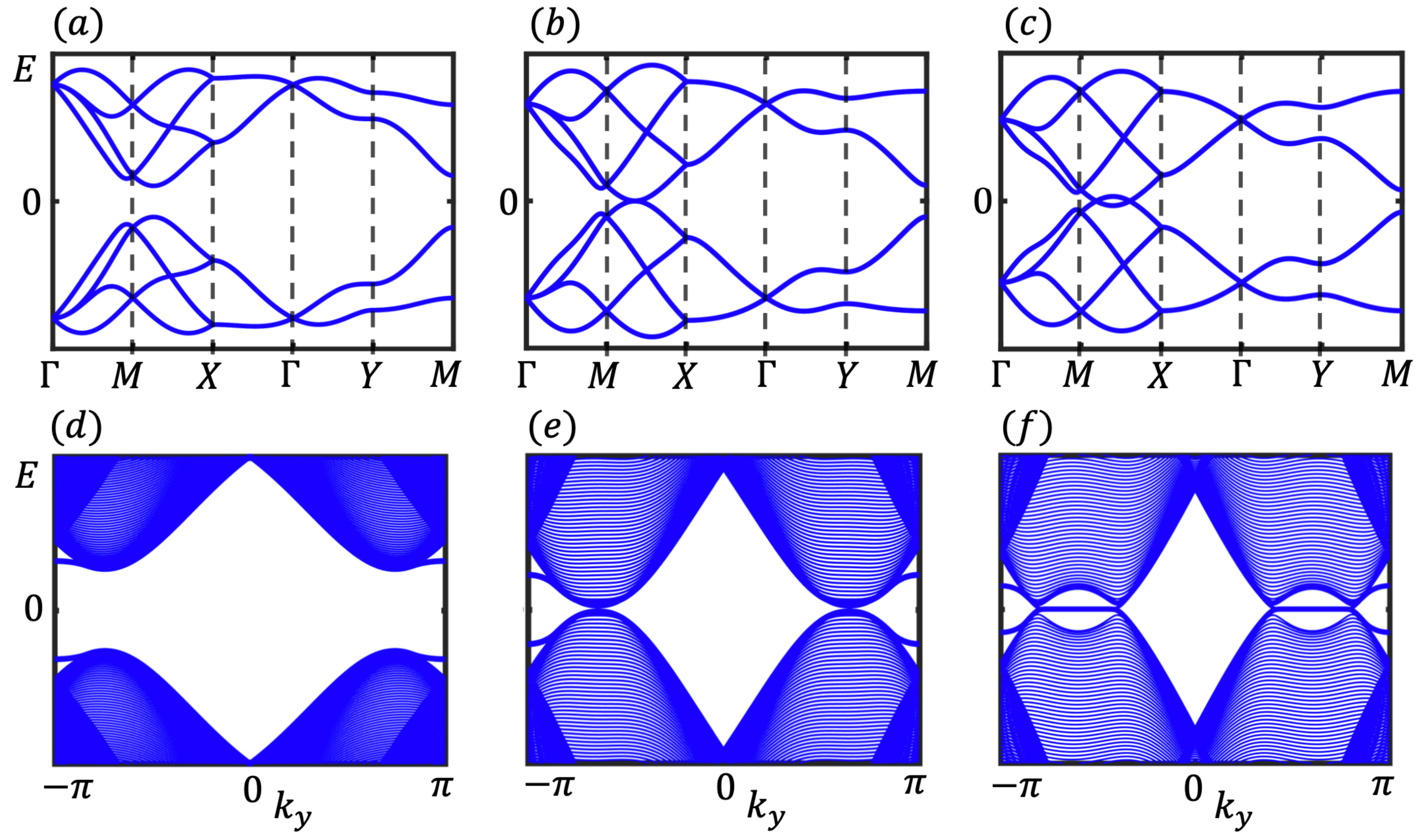}
\caption{Upper: The band structure of system with nearest-neighbor coupling when $\lambda/\gamma=0.4$ for different coupling amplitudes. (a) $\Omega/\gamma=0.3$, (b) $\Omega/\gamma=0.48$, (c) $\Omega/\gamma=0.6$. Lower: The band structure with open boundary along $x$ direction with corresponding system parameters. }\label{fig:bandchiralcoupling}
\end{figure}

It is easy to verify that the system still has the glide reflection symmetry with the same glide operator as the above discussion.The chiral symmetry and $C_2$ symmetry are also preserved with corresponding symmetric operators as $\mathcal{C}=\tau_z\sigma_z$ and $C_2=\tau_x\sigma_y$, which means that the system is still a second-order topological insulator. The band structure is shown in fig.\ref{fig:bandchiralcoupling}. This type of coupling also preserve the reflection symmetries along $x$ and $y$ directions with the corresponding symmetric operators change into $M_x=s_z\sigma_x$ and $M_y=s_z\tau_x\sigma_z$ respectively.
In most area of the BZ, two-fold degeneracy has been lifted. But along some high-symmetric lines, such as along $k_x=0$ and $k_y=0,\pi$ lines, the band structure still has two-fold degeneracy. We can define another operator $\Theta'=is_z\tau_x\sigma_y\mathcal{K}$ which is commute with original Hamiltonian (\ref{hotiwithglideinmomentumspace}) in Section \ref{sec2} when $\Omega=0$. It also commute with those $k_x$ dependent terms in the coupling Hamiltonian (\ref{hotiwithchiralcoupling}). So along $k_y=0$ line, $\Theta'$ is commute with the Hamiltonian of the whole system and the band has Kramers degeneracy. Along $k_x=0$ line, the operator $\Theta=i\tau_x\sigma_y\mathcal{K}$ can be used to explain the degeneracy. Along $k_y=\pi$ line, the Kramers degeneracy can not be explained easily because the corresponding symmetric operator should be ${\bf k}$-dependent.

The most interesting phenomena is along $k_x=\pi$, where the two-fold degeneracy is lifted. With the increase of coupling amplitude, there are two band crossing points with quadratic dispersions along this line at critical point and the system has a topological phase transition, as shown in fig.\ref{fig:bandchiralcoupling}(b) (it only shows the $k_y>0$ one). After passing through the critical point, these two band crossing points separate into four band crossing points with linear dispersions and the system becomes a semimetal, as shown in fig.\ref{fig:bandchiralcoupling}(c). To explain these transition analytically, one can consider an ${\bf k}$-dependent unitary transformation as $U_{\bf k}=\mathrm{diag}\big[\mathcal{I}_{4\times 4},-ie^{-ik_y/2}\tilde{G}_{\bf k}\big]$ where $H'_{\bf k}=U^\dag_{\bf k}H_{k}U_{\bf k}$. With this transformation, from relation (\ref{blockglideops}), two diagonal blocks of $H'_{\bf k}$ become the same and the form of whole Hamiltonian along $k_x=\pi$ line can be written as
\begin{equation}
\begin{split}
H'_{k_x=\pi}&=(\lambda-\gamma)\tau_z\sigma_x+(\lambda+\gamma\cos k_y)\tau_x+\gamma\sin k_y\tau_y \\
-&2\Omega\sin\frac{k_y}{2}s_x\sigma_x+2\Omega\cos\frac{k_y}{2}s_x\tau_x+2\Omega\sin\frac{k_y}{2}s_x\tau_y
\label{kypichiralcoupling}
\end{split}
\end{equation}
which can be diagonalized analytically to get the energy spectrum as
\begin{equation}
E_{k_x=\pi}=\pm 2\Omega\sin\frac{k_y}{2}\pm\sqrt{(\lambda-\gamma)^2+\Big|\lambda+\gamma e^{ik_y}\pm 2\Omega e^{\frac{ik_y}{2}}\Big|^2}
\label{espectrumchiralcoupling}
\end{equation}
Then the critical point can be estimated as
\begin{equation}
\Omega_\mathrm{cri}=\sqrt{\frac{2\lambda\gamma(\lambda-\gamma)^2}{6\lambda\gamma-\lambda^2-\gamma^2}}
\label{cripointchiralcoupling}
\end{equation}
We can also choose the open boundary along $x$ direction. As shown in fig.\ref{fig:bandchiralcoupling}(f), the zero-energy edge states emerge in the restricted region between two band crossing points in the semimetal phase, which is similar with the bearded and zigzag edges of graphene\cite{graphene}. An interesting property is that the relation (\ref{cripointchiralcoupling}) is meaningful only when the value under the square root is positive, which means that the topological phase transition can only happen when $(3-2\sqrt{2})\gamma<\lambda<\gamma$. If $\lambda\ll\gamma$, there is no topological phase transition and the system is always a non-trivial HOTI with the increase of coupling amplitude. Some details of this case is discussed in Appendix \ref{appsec2}.

\begin{figure}[tph]
\centering
\includegraphics[width=80mm]{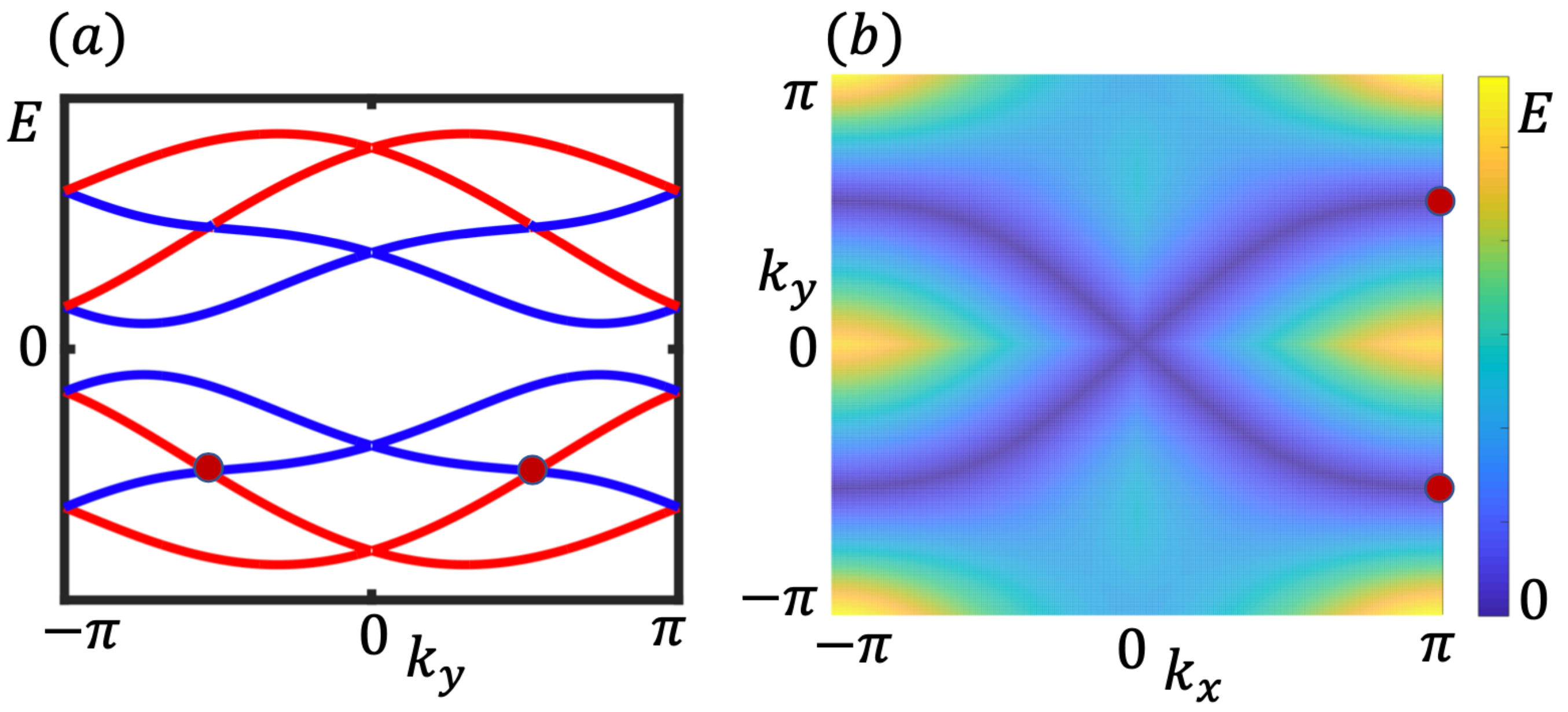}
\caption{(a) The band structure for $k_x=\pi$. Red and blue lines correspond to the sign as $+$ and $-$ before the eigenvalue of glide operator. (b) The gap between the second and third bands. The color from dark to light correspond to the increase of gap energy. The darkest part corresponds to the band crossing points. Two red dots in (a) and (b) correspond to each other. The tunneling and coupling amplitude we choose in (a) and (b) are $\lambda/\gamma=0.4$ and $\Omega/\gamma=0.3$ respectively.}\label{fig:bandcrossingforky}
\end{figure}
 
In most area of BZ, Kramers degeneracy is lifted and glide reflection symmetry induces additional band crossing. As shown in fig.\ref{fig:bandcrossingforky}(a), different bands correspond to different eigenvalues of the glide operator. Red and blue lines in fig.\ref{fig:bandcrossingforky}(a) correspond to the sign as $+$ and $-$ before the eigenvalues of glide operator respectively. For an arbitrary $k_y$, $e^{i(k_x+k_y)/2}$ will return to the original value after a transition $k_y\rightarrow k_y+4\pi$, which passing through twice of BZ. So there should be at least one band crossing point in the first BZ, as the red dots in fig.\ref{fig:bandcrossingforky}. Furthermore, the band crossing points protected by the glide reflection symmetry is robust to the change of system parameters such as coupling amplitude. As shown in fig.\ref{fig:bandcrossingforky}(b), the band crossing points between the second and third bands come into a nodal line whose form has no qualitative changes in distinct topological phases.

\section{Magnetic glide symmetry}
\label{sec4}

In this section, we consider another type of on-site coupling with the corresponding coupling Hamiltonian as
\begin{equation}
\begin{split}
H_\mathrm{couple}=&2\Omega\sum_{mn}\big(e^{i\varphi}a^\dag_{mn,\uparrow}a_{mn,\downarrow}+e^{-i\varphi}b^\dag_{mn,\uparrow}b_{mn,\downarrow} \\
&+e^{-i\varphi}c^\dag_{mn,\uparrow}c_{mn,\downarrow}+e^{i\varphi}d^\dag_{mn,\uparrow}d_{mn,\downarrow}\big)+h.c.
\end{split}
\end{equation}
where $\varphi$ is an additional phase in the coupling. With the Fourier transformation, it can be expressed in momentum space as
\begin{equation}
H_\mathrm{couple}=2\Omega\cos\varphi s_x+2\Omega\sin\varphi s_y\tau_z\sigma_z
\label{magneticglidecoupling}
\end{equation}
When $\varphi=0,\pi/2,\pi$, this coupling term is equivalent to the case we discussed in Section \ref{sec2}, which preserve both the glide reflection symmetry and time-reversal symmetry. But when $0<\varphi<\pi/2$ or $\pi<\varphi<\pi$, both of these two symmetries has been broken, only their combination $G_{\bf k}\mathcal{T}$ is preserved. This symmetry is one of the nonsymmorphic magnetic symmetries\cite{glide3} and we call it magnetic glide symmetry. We can also find that the system still has the chiral symmetry and $C_2$ symmetry with the same symmetric operators as in Section \ref{sec2}, which means that the system is still a HOTI. This type of coupling break the reflection symmetries along $x$ and $y$ directions, but with unitary operators $\tilde{M}_x=\tau_x\sigma_z$ and $\tilde{M}_y=\sigma_x$, we have the relations $\tilde{M}^{-1}_xH_{(-k_x,k_y)}\tilde{M}_x=H^*_{(k_x,k_y)}$ and $\tilde{M}^{-1}_yH_{(k_x,-k_y)}\tilde{M}_y=H^*_{(k_x,k_y)}$ respectively, where $H^*_{(k_x,k_y)}$ corresponds to the Hermitian conjugation of Hamiltonian $H_{(k_x,k_y)}$. So the band structure is still symmetric along $k_x$ and $k_y$ directions.

\begin{figure}[tph]
\centering
\includegraphics[width=87mm]{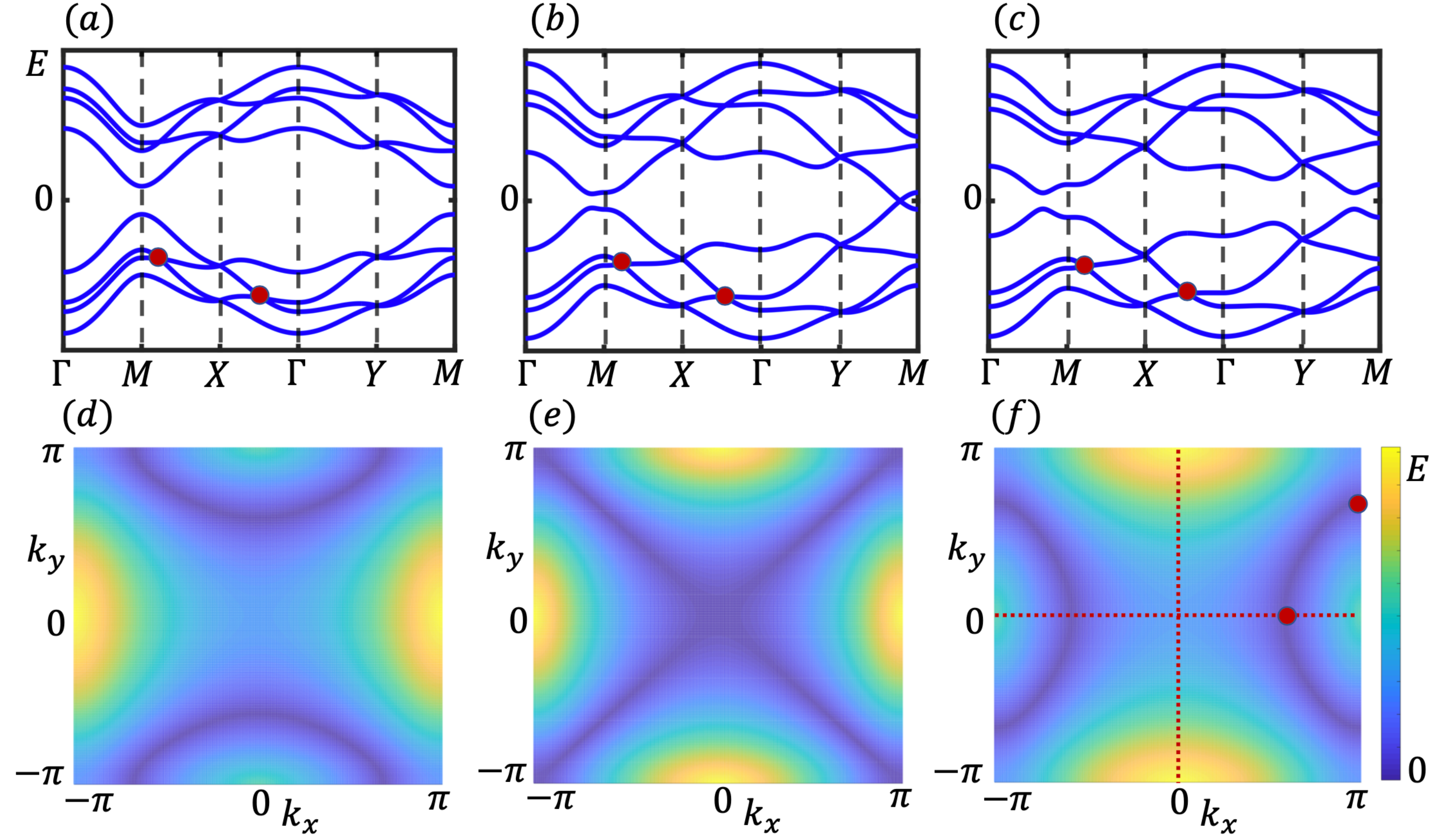}
\caption{(a)-(c) correspond to the band structure when $\lambda/\gamma=0.4$ with different coupling amplitude $\Omega$ where $\varphi=\pi/3$. (a) $\Omega/\gamma=0.3$, (b) $\Omega/\gamma=0.48$, (c) $\Omega/\gamma=0.6$. The red dots correspond to the band crossing points protected by the magnetic glide symmetry. (d)-(e) corresponds to the gap between the second and third bands with different $\varphi$ when the coupling amplitude is fixed as $\Omega/\gamma=0.3$. The color from dark to light correspond to the increase of gap energy. The darkest part corresponds to the band crossing points. (d) $\varphi=\pi/6$, (e) $\varphi=\pi/4$, (f) $\varphi=\pi/3$. The red dots in (f) correspond to the positions of two band crossing points along $k_x=\pi$ and $k_y=0$ lines in BZ. }\label{fig:magneticglide}
\end{figure}

The band structure is shown in fig.\ref{fig:magneticglide}(a)-(c). With the increase of coupling amplitude, the system has three different topological phases, which is very different with above two cases. Numerical calculation shows that there are two critical points in the system as $\Omega_{1,\mathrm{cri}}$ and $\Omega_{2,\mathrm{cri}}$ which depend on the value of $\varphi$ and have no explicit expressions. When the coupling amplitude $\Omega<\Omega_{1,\mathrm{cri}}$, the system is a non-trivial HOTI. When $\Omega=\Omega_{1,\mathrm{cri}}$, the system translates into a semimetal with two band crossing points with quadrupole dispersion. After passing through this critical point, these two band crossing points separate into four band crossing points with linear dispersions and moving in the first BZ with the increase of $\Omega$. When $\Omega$ approaches to $\Omega_{2,\mathrm{cri}}$, the four band crossing points merge into two crossing points again. After passing through $\Omega_{2,\mathrm{cri}}$, band gap will be reopened and the system becomes topological trivial. At two sides when $\Omega<\Omega_{1,\mathrm{cri}}$ and $\Omega>\Omega_{2,\mathrm{cri}}$, the topology of bulk and corner are similar with the case we discussed in Section \ref{sec2}. 

Because the breaking of time-reversal symmetry, two-fold degeneracy has been lifted in most area of BZ. But the magnetic glide symmetry also lead to additional band crossing points between the second and third bands, which forms a nodal ring. As shown in fig.\ref{fig:magneticglide} (d)-(f), the form of nodal ring has no qualitative changes in different topological phases with the increase of $\Omega$. It only depends on the value of the phase of coupling $\varphi$.

\section{Scheme of experimental setup}
\label{sec5}

Exploring the models we discussed above in condensed matter systems directly requires more in-depth research. But the synthetic quantum systems provide convenient platforms for simulating novel topological states. For example, we can try to simulate the coupled-BBH model in ultracold atomic systems with the following three steps. The first step is simulating BBH model using two-dimensional superlattice. The second step is constructing a spin-dependent lattice where lattice structure of different spin component has a displacement along particular direction. The third step is adding the coupling using microwave or laser fields.

\begin{figure}[tph]
\centering
\includegraphics[width=86mm]{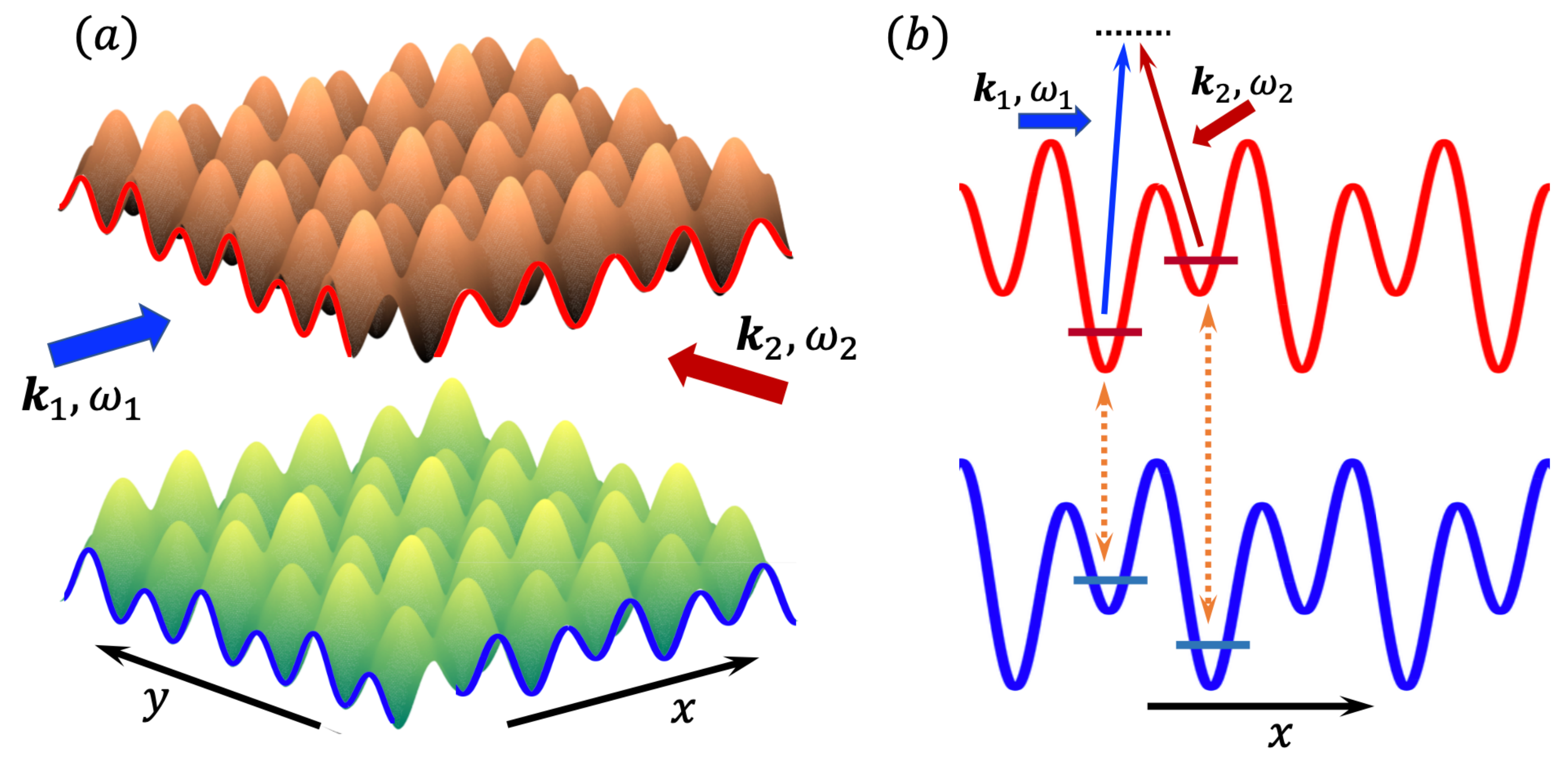}
\caption{Scheme of experimental setup. (a) The lattice structure for different spin components. The spin-up and spin-down are shown by upper and lower surfaces respectively. The red and blue lines correspond to the lattice potential along $x$ and $y$ directions for different spin components respectively. Red and blue arrows are the Raman lasers with wave vectors and frequencies as ${\bf k}_1, \omega_1$ and ${\bf k}_2, \omega_2$ respectively. (b) The lattice structure along $x$ direction for spin-up (upper) and spin-down (lower) respectively. The photon-assisted nearest-neighbor tunneling is induced by Raman process. Two dotted lines with double arrow correspond to the microwave fields coupling different spin components on different types of sublattices respectively. }\label{fig:latticestructure}
\end{figure}

{\it Step 1}. Consider the ultracold atoms confined in two-dimensional superlattice, which is constructed by two types of counter-propagating lasers with long and short wavelength $4\pi/k_0$ and $2\pi/k_0$ respectively ($k_0$ is the wave vector of laser with short wavelength) along both $x$ and $y$ directions, as shown in fig.\ref{fig:latticestructure}. The lattice structure can be written as $V(x,y)=V(x)+V(y)$ where
\begin{equation}
\begin{split}
V(x)=&V_{sx}\sin^2(k_0x)+V_{lx}\sin(k_0x+\theta) \\
V(y)=&V_{sy}\sin^2(k_0y)+V_{ly}\sin(k_0y)
\label{latticepot1}
\end{split}
\end{equation}
where $0<\theta<\pi/2$ is the relative phase between two lasers with different wavelength propagating along $x$ direction. The unit cell of the two-dimensional superlattice is a square which including four neighbor sites. Along $y$ direction, the difference between intra- and inter-cell nearest-neighbor tunneling lead to a Su-Schrieffer-Heeger model\cite{ssh, zakphase}. But as shown in fig.\ref{fig:latticestructure}(b), the lattice structure along $x$ direction is asymmetric because of the non-zero relative phase $\theta$, the nearest-neighbor tunneling along $x$ direction is forbidden by the energy detuning between two neighbor sites. By adding two additional Raman lasers with the frequency difference matching this energy detuning exactly, the Raman process leads to a two-photon assisted tunneling with an additional phase, which corresponding to adding an effective staggered flux in the superlattice\cite{fluxlattice}. The asymmetric of the lattice structure also induce the difference between intracell and intercell tunneling amplitude. Setting the wavelength of two Raman lasers equal to the wave length of short laser, the flux can be set as $\pi$, which corresponding to the BBH model. The amplitude of tunneling along $x$ direction can be fine tuned by choosing the Rabi frequencies of Raman lasers. But we should point out that the difference of tunneling amplitudes along $x$ and $y$ directions will not break the symmetries discussed in our work.

{\it Step 2}. There are a lot of mature experimental technics to realize the construction of spin-dependent lattices in ultracold atomic systems\cite{spinlattice1, spinlattice2, spinlattice3}. We can also use two types of lasers with the same wave vector but different frequencies to confine two different hyperfine spin states of the same type of atoms individually. If the lattice potential of spin-up component is (\ref{latticepot1}), the lattice potential of spin-down component can be written as $V'(x,y)=V'(x)+V'(y)$ where $V'(x)=V_{sx}\sin^2(k_0x)-V_{lx}\sin(k_0x+\theta)$ and $V'(y)=V_{sy}\sin^2(k_0y)-V_{ly}\sin(k_0y)$, where the lattice potential with long wavelength has an opposite sign. It is the same with the lattice potential of spin-up but a shift of half lattice spacing along both $x$ and $y$ direction, which can be realized by a phase shift between these two types of lasers. Then with the same Raman process on two different types of superlattices, the lattice structure shown in fig.\ref{fig:latticeonsitecoupling} can be constructed.

{\it Step 3}.  We have to point out that it is not easy to construct the type of coupling as discussed in Section \ref{sec3}. Because it is not a Rashba-type spin-orbit coupling\cite{soc2d1, soc2d2, soc2d3} but a nearest-neighbor coupling with additional phase along clockwise ($a_\uparrow, c_\uparrow$) and anti-clockwise ($d_{\uparrow}, b_\uparrow$) directions respectively. But the realization of on-site coupling discussed in Section \ref{sec2} and \ref{sec4} is not difficult. As shown in fig.\ref{fig:latticestructure}(b), because the asymmetric lattice potential along $x$ direction, the energy detuning between different spin components on $a(c)$ and $b(d)$ sublattices are different. So two microwave fields with different frequencies need to be added to couple different spin components on $a(c)$ and $b(d)$ sublattices respectively. The frequencies of two microwave fields should match the energy detuning of two different types of sublattices respectively. By locking the relative phase as zero between these two microwave fields, the on-site coupling in Section \ref{sec2} can be realized. 

The realization of the on-site coupling in Section \ref{sec4} is more complex. We need to construct the coupling with the form as $\Omega_1s_y\pm\Omega_2\cos(k_0x)s_x$ where $\varphi\approx\tan^{-1}(\Omega_2/\Omega_1)$, which can be separated into two parts. The first constant part with the form $\Omega_1s_y$ can be realized by adding microwave fields as discussed above. To realize the spatial-modulated coupling $\Omega_2\cos(k_0x)s_x$, the microwave fields may not work, but the Raman process can be used to instead. We can add two counter-propagating Raman lasers with frequency $\omega_3$ to constitute the spatial-modulated field as $\propto\cos(k_0x)e^{-i\omega_3t}$. If the frequency is fine tuned as $\omega_3-\omega_1$ is matching the energy detuning between different spin components on $a(c)$ sublattices, the $\Omega_2\cos(k_0x)s_x$ form of coupling can be realized. The spatial-modulated coupling on $b(d)$ sublattices can be constituted with another two counter-propagating Raman lasers with frequency $\omega_4$ where $\omega_4-\omega_1$ is matching the energy detuning on $b(d)$ sublattices. A relative phase is necessary to these two Raman lasers to add an opposite sign before the spatial-modulated coupling on $b(d)$ sublattices comparing with $a(c)$ sublattices.

\section{Discussion}
\label{sec6}

In this work, we propose a coupled-BBH model with different types of coupling and explore the interplay between higher-order topology and glide reflection symmetry, where interesting band crossings and degeneracies are predicted. We have also discussed the corresponding experimental implementation methods of two schemes in ultracold atomic systems. The band crossing points and nodal rings discussed in our work can be observed with band mapping technics\cite{bandmapping}. In principle, there is no difficulties to construct these systems and observe the corresponding properties in other synthetic quantum systems and metamaterials. 

In the future, we will extend our exploration to higher-dimensional system. The influence of nonsymmorphic symmetries to the three-dimensional higher-order topological systems and their surface states should be more interesting and have more potential applications than the present work. Another interesting research in the future is the generalization of the present model into many-body systems by considering the interaction between different spin components. We hope that the interplay between higher-order topology, nonsymmorphic symmetries and interaction will lead to more interesting quantum phenomena.

\section*{Acknowledgements}

We acknowledge the useful discussion with Xi-Wang Luo. The work is supported by the National Natural Science Foundation of China (Grant No. 12174138).

\begin{appendix}

\section{The topological properties of system in Section \ref{sec3} when $\lambda\ll\gamma$}
\label{appsec2}

\begin{figure}[tph]
\centering
\includegraphics[width=86mm]{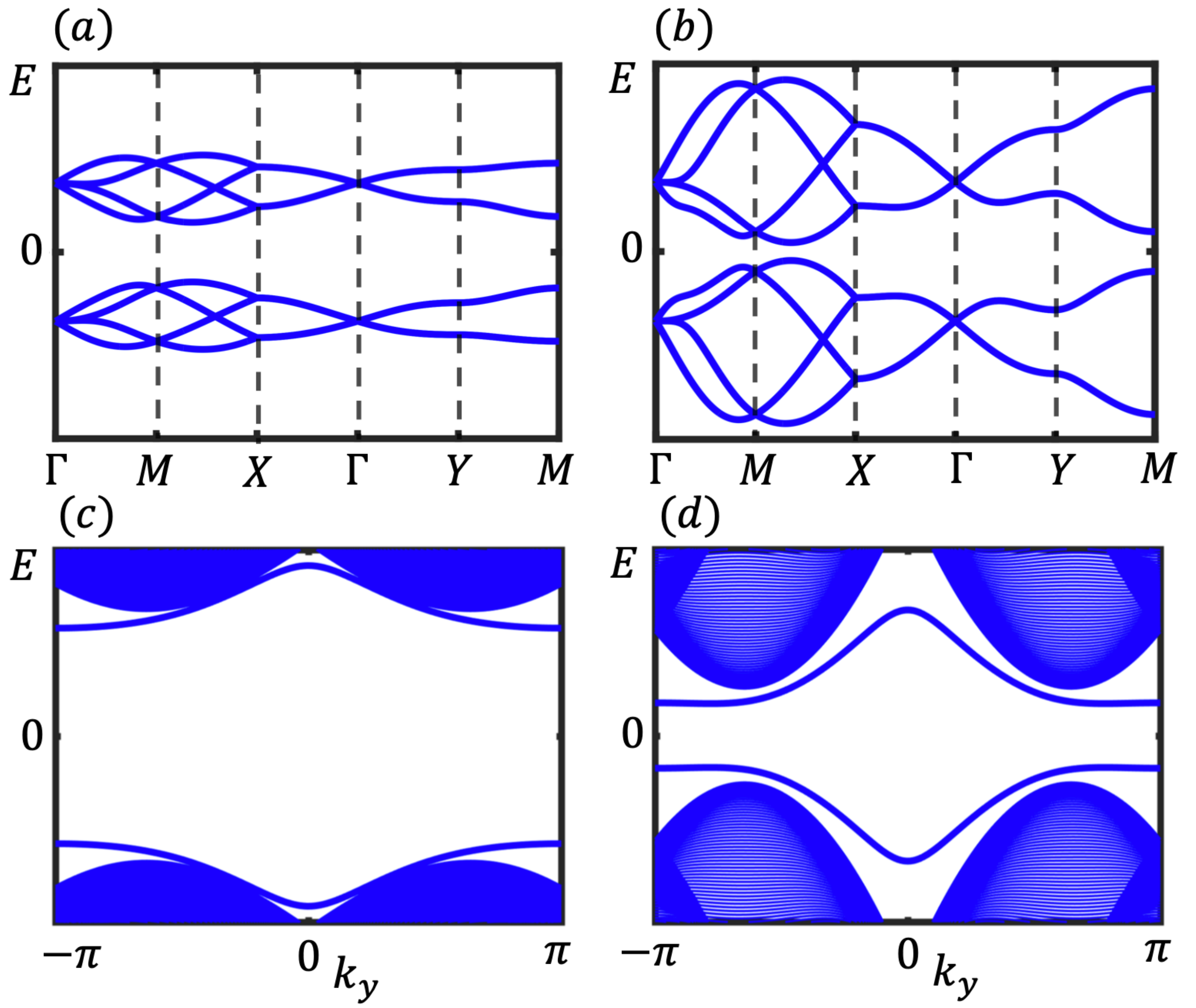}
\caption{Upper: The band structure of system with nearest-neighbor coupling when $\lambda/\gamma=0.1$ for different coupling amplitudes. (a) $\Omega/\gamma=0.3$, (b) $\Omega/\gamma=0.8$. Lower: The band structure with open boundary along $x$ direction with corresponding system parameters. }\label{fig:edgenotransition}
\end{figure}

When $\lambda\ll\gamma$, the system with coupling Hamiltonian (\ref{hotiwithchiralcoupling}) has no topological phase transition. With the increase of coupling amplitude, the system is always HOTI with four zero-energy corner states. As shown in fig.\ref{fig:edgenotransition}, with the increase of coupling amplitude, the band width increase and the energy of middle two bands along $k_x=\pi$ will approach to zero. But they will never touch the zero energy line. From the energy spectrum (\ref{espectrumchiralcoupling}), the position of gap minimum can be estimated as 
\begin{equation}
k_y=2\cos^{-1}\frac{\Omega(\lambda+\gamma)}{2(\Omega^2+\lambda\gamma)}
\end{equation}
which approaching to $\pi$ when $\Omega$ is far more larger than $\lambda$ and $\gamma$. The corresponding minimum gap is about $\Delta E\approx\frac{(\lambda+\gamma)^2\Omega^3}{4(\Omega^2+\lambda\gamma)^2}$ which approaching to zero.

With open boundary along $x$ direction, the band structure is different with the case we discuss in Section \ref{sec3}. There are two separated bands which corresponding to edge states exist in the middle of the bulk energy bands. With the increase of coupling amplitude, these two edge bands move approach to zero energy line, but they will also never touch it.

To show the topological properties of the system. We can consider the extreme case where $\lambda=0$. In this situation, four zero-energy corner states are all localized on the diagonal lines of the lattice and have the analytical form. One of them can be written as
\begin{equation}
\begin{split}
&|\Psi\rangle_\mathrm{corner}\propto \\
&\sum_nx^{2(n-1)}\big(a^\dag_{nn,\uparrow}+ixa^\dag_{nn,\downarrow}+xd^\dag_{nn,\downarrow}+ix^2d^\dag_{nn,\uparrow}\big)|0\rangle
\end{split}
\end{equation}
where $|0\rangle$ is the vacuum state and
\begin{equation}
x=\frac{(1+i)}{2\Omega}\big(\lambda-\sqrt{2\Omega^2+\lambda^2}\big)
\end{equation}
It is easy to verify that state $|\Psi\rangle_\mathrm{corner}$ is the eigenstate with zero energy. One can also easily verify that $|x|<1$ is satisfied in the whole range of $\Omega$, which means that this state is localized on the corner of lattice and is one of the zero-energy corner states.

\end{appendix}

\end{document}